\documentclass[useAMS,usenatbib]{mn2e}
\usepackage{graphicx}                                            
\usepackage{times}
\usepackage{epsfig}
\usepackage{rotating}

\newcommand{\ha}{H$\alpha$}

\def\kms{\mbox{${\rm km}\:{\rm s}^{-1}\:$}}

\def\lesssim{\mathrel{\hbox{\rlap{\hbox{\lower4pt\hbox{$\sim$}}}\hbox{$<$}}}}
\let\la=\lesssim
\def\gtrsim{\mathrel{\hbox{\rlap{\hbox{\lower4pt\hbox{$\sim$}}}\hbox{$>$}}}}
\let\ga=\gtrsim

\def\sol{~\mathrm{M}_\odot}

\title[The black hole candidate XTE J1752-223]{The hard state of black hole candidates: XTE J1752-223 }
\author[Mu\~noz-Darias et al.]{T.~Mu\~noz-Darias$^{1}$, S.~Motta$^{1,2}$, D.~Pawar$^{3}$, T.~M.~Belloni$^{1}$, S.~Campana$^{1}$ and D.~Bhattacharya$^{4}$ \\
$^{1}$INAF-Osservatorio Astronomico di Brera, Via E. Bianchi 46, I-23807 Merate (LC), Italy\\
$^{2}$Universit\`a dell'Insubria, Via Valleggio 11, I-22100 Como, Italy \\
$^{3}$Department of Physics, Ramniranjan Jhunjhunwala College, Ghatkopar (W), Mumbai 400 086, India\\
$^{4}$Inter-University Center for Astronomy and Astrophysics (IUCAA), Post bag No. 4, Ganeshkhind, Pune 411007, India}

\begin{document}
\maketitle
\begin{abstract}
We present a two-day long RXTE observation and simultaneous Swift data of the bright X-ray transient XTE J1752-223. Spectral and timing properties were stable during the observation. The energy spectrum is well described by a broken power-law with a high energy cut-off. A cold disc ($\sim 0.3$ keV) is observed when Swift/XRT data are considered. The fractional rms amplitude of the aperiodic variability (0.002--128 Hz)  is $48.2\pm0.1\%$ and it is not energy dependent. The continuum of the power density spectrum can be fitted by using four broad-band Lorentzians. A high frequency ($\sim 21$ Hz) component and two weak QPO-like features are also present. Time-lags between soft and hard X-rays roughly follow the relation $\Delta t \propto \nu^{-0.7}$, with delays dropping from $\sim 0.5$ (0.003 Hz) to $\sim 0.0015$ ($\geq$10 Hz) seconds. Our results are consistent with XTE J1752-223 being a black-hole candidate, with all timing and spectral components very similar to those of Cyg X-1 during its canonical hard state.
\end{abstract}
\section{Introduction}
Black hole X-ray transients (BHT) represent the  majority of the black hole binary (BHB) population known so far. These systems spend most of their lives in quiescence, displaying luminosities too low to be detected by X-ray all-sky monitors (see e.g. \citealt{Garcia1998}).  However, they also undergo outburst events in which they become as bright as persistent sources, allowing their discovery. During these episodes, both the spectral and the time variability properties of BHTs vary dramatically, yielding the so-called \textquoteleft states\textquoteright. There is still much discussion about how many different states there are, and their correspondence with different physical conditions (see e.g. \citealt{Belloni2010} for a general description), but the presence of a $hard$ state (historically known as \textit{low/hard}; LHS) at the begining of the outburst which evolves towards a $soft$ state (\textit{high/soft}; HSS) is widely accepted. The LHS, also associated with the last part of the outburst, is characterized by a power-law dominated energy spectrum with a power-law index of $\sim 1.6$ (2--20 keV band). A high energy cut-off ($\sim$ 60--200 keV; \citealt{Wilms2006}; \citealt{Motta2009}) is observed and aperiodic variability with a fractional root mean square amplitude (rms) above 30\% is seen. The energy spectrum is softer during HSS, being dominated by a thermal disc black body component. However, a hard tail up to $\sim 1$ MeV is present (\citealt{Grove1998}). The rms associated with the aperiodic variability drops until 1\% or less. These two \textquoteleft canonical\textquoteright ~states were first proposed to describe the behaviour of the prototypical BHB Cyg X-1.\\
XTE J1752-223 was discovered by the \textit{Rossi X-ray Timing Explorer} (RXTE) on October 23, 2009 (\citealt{Markwardt2009}). The source showed a 2--10 keV flux of 30 mCrab. Significant similarities with the typical properties of BHT during the LHS were soon noticed by \cite{Markwardt2009a} and \cite{Shaposhnikov2009}. A bright optical counterpart was detected (\citealt{Torres2009}), showing in the optical spectrum a broad \ha~emission line ($FWHM\sim750$ \kms) typical of accreting binaries (\citealt{Torres2009a}). A radio counterpart with a spectrum consistent with that of a compact jet, as expected for LHS, was also reported by \cite{Brocksopp2009}. All these properties triggered a daily monitoring by RXTE in order to follow up the outburst evolution. In this paper we present spectral and time variability analysis of XTE J1752-223 using an unusually long, almost continuous observation ($\sim 116$ ks) performed by RXTE during 26th, 27th and 28th October 2009 (MJD 55130-55132). The quality of this data set  allows us to perform a detailed study of this source and compare its general behaviour with that shown by the prototypical BHB Cyg X-1 during LHS. 

\begin{figure}
\centering
 \includegraphics[width= 8.8cm,height=6.5cm]{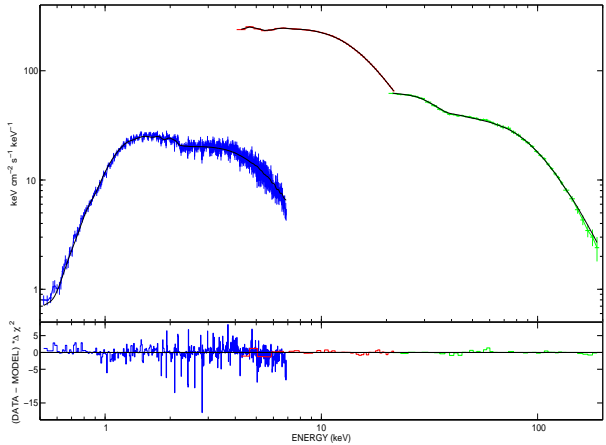}
\caption{Our best fit to the combined XRT, PCA and HEXTE spectrum of XTE J1752-223. The model used consist of an interstellar absorption component, two Gaussian lines, a multicolour black-body disc, a broken power-law and a high energy cut-off (see text). Top panel: XRT, PCA and HEXTE spectra. Bottom panel: residuals plotted as signed contribution to the chi square.}
\label{fig:1752_sp}
\end{figure}

\section{Observations}
We analyse a  $\sim 116$ ks observation of XTE J1752-223 interrupted only by satellite-related gaps. This corresponds to RXTE archive observations identifiers 94044--07--01--00, 94044--07--01--01, 94044--07--01--02 and 94331--01--01--00, 94331--01--01--02 that were performed between October 26 (15:03 UT) and 29 (20:50 UT), 2009. A total of 35 one-orbit pointings were used. For comparison we have also analysed  a $\sim 15$ ks RXTE observation of Cyg X-1 (94121--01--08--00) starting on April 12, 2009 (02:15 UT). This observation was chosen because it is the longest observation of Cyg X-1 performed recently with RXTE and the instrument response is expected to be similar to that for the XTE J1752-223 data.\\ 
The variability study presented in this paper is based on data from the \textit{Proportional Counter Array} (PCA). For XTE J1752-223 the data are in the mode E\_125us\_64M\_0\_1s which covers the PCA effective energy range (2-60 keV) with 64 bands. However only PCA channels 0--35 (2--15 keV) were used in the analysis. For Cyg X-1 we selected the modes SB\_125us\_0\_13\_1s and SB\_125us\_0\_13\_1s that cover the 2--15 keV band used for the analysis of  XTE J1752-223. For both objects the time resolution is 122$\mu$s.\\ 
The PCA Standard 2 mode (STD2) was used for spectral analysis. It covers the 2--60 keV energy range with 129 channels. From the data, we extracted hardness ($h$), defined as the ratio of counts in STD2 channels 11--20 (6.1--10.2 keV) and 4-10 (3.3--6.1 keV). Using this definition we find similar average values of $h=0.887\pm0.002$ for XTE J1752-223 and  $h=0.821 \pm 0.003$ for Cyg X-1, both in the range expected for LHS. 
Energy spectra from the PCA and \textit{High Energy X-ray Timing Experiment} (HEXTE) instruments (background and dead-time corrected) were extracted for each observation using the standard RXTE software within \textsc{heasoft} V. 6.7. For our spectral fitting, only Proportional Counter Unit  2 from the PCA and Cluster B from HEXTE were used. In order to account for residual uncertainties in the instrument calibration a systematic error of $0.6\%$ and $1\%$  was added to the PCA and HEXTE spectra, respectively. All the observations were averaged in a single spectrum after a preliminary spectral analysis where no significant spectral variability across the pointings was found.\\ 
To extend our energy coverage down to 0.5 keV we also made use of snapshot \textit{Swift X-ray Telescope} (XRT) observations collected within 26-28 Oct 2009. In particular, we use observation numbers 00031532001, 00031532002 and 00031532003 (3.2 ks of data). They were carried out in Windowed Timing (WT) mode, due to the brightness of the source. In WT mode a 1D image is obtained by reading data compressed along the central 200 pixels in a single row. The XRT data were reprocessed with standard procedures (xrtpipeline V. 0.12.3 within \textsc{heasoft} V. 6.7). We extracted source events in a circular region with radius 20 pixels centred on source. Ancillary response files were generated with the {\sc xrtmkarf} task, accounting for CCD defects, vignetting, and point-spread function corrections. Given the relatively high interstellar absorption and source count rate, we prefer also to select photons based on their grade, allowing only for single-pixel events. This provides a clearer spectrum and a higher spectral resolution. The source is very bright: since within the energy range considered the background contributes less than 1\%, we did not correct for the background. After verification of comparable spectra in the three observations, we summed the data into a single spectrum, creating the corresponding \textit{arf} file. During the fit we used the response file swxwt0s6\_20070901v011.rmf appropriate for single pixel events and for the new XRT substrate voltage (6 V).

\begin{table}
\begin{center}
\begin{tabular}{|c|c|}
\hline
Spectral parameter		&    		Value		\\
\hline
\hline
Absorption ($10^{22}$ atoms cm$^{-2}$) 			&      	$0.72_{-0.04}^{+0.01}$ \\
$T_\mathrm{in}$ (keV)			&      	$0.313 \pm 0.007$ \\
\rm{Diskbb norm.}		&     ($1.027\pm0.001$) $\times 10^6$	 \\
$\Gamma_{1}$		&     $1.471 \pm 0.008$	 \\			
$E_\mathrm{break}$	(keV)		&   $10.2 \pm 0.4$ \\
$\Gamma_{2}$		&  $1.24 \pm 0.01$ \\
\rm{PL. norm.} (photons keV$^{-1}$cm$^{-2}$s$^{-1}$) 			&      $44.7_{-0.6}^{+0.5}$	  \\			
High energy cut-off (keV) 		&      	$133_{-5}^{+6}$  \\
\hline
\end{tabular}
\caption{
Spectral parameters for XRT+PCA+HEXTE spectra. Absorption is the equivalent Hydrogen column. $T_\mathrm{in}$ is the temperature at inner disc radius. \rm{Diskbb norm.}~is the normalisation of the \textit{diskbb} component defined as $(\frac{R_{in}/ \mathrm{km}}{D/10~\mathrm{kpc}})^2 cos \Theta$,  where $R_\mathrm{in}$ is the inner disc radius (km), $D$ is the distance to the source (kpc) and $\Theta$ is the inclination angle of the disc. $\Gamma_{1}$ and $\Gamma_{2}$ are the power-law photon indexes. $E_{\mathrm{break}}$~is the break energy. \rm{PL norm.} is the power-law normalisation at 1 keV.}
\label{tab:spectral_par}
\end{center}
\end{table}

\section{Data analysis}
The analysis of the energy spectrum, power density spectrum (PDS), rms spectrum and time-lags is presented here. They have been performed making use of \textsc{xspec} and custom timing software running under \textsc{idl}.  
\subsection{Spectral analysis}
In order to perform a broad band spectral analysis, \textsl{Swift}/XRT (0.5-7.0 keV), PCA (4-20 keV) and HEXTE (20-200 keV) spectra were combined. \textsc{xspec} V. 12.5.1 was used to fit the spectra. We started by trying a one-component model, either a cut-off-power-law (\textit{cutoffpl} in \textsc{xspec}) or a black body disc (\textit{diskbb} in \textsc{xspec}), without success. In the same way, a clear residual in the soft part of the spectrum is observed when using a simple combination of these two models (i.e. \textit{cutoffpl} + \textit{diskbb}).\\
\cite{Nowak2005} showed that the energy spectra of BHB can be empirically described by an absorbed broken power-law with an exponential cut-off. Following this, we substituted the power-law component by a broken power-law component (\textit{bknpower} in \textsc{xspec}). We also found that a high energy cut-off component (\textit{highecut} in \textsc{xspec}) with a typical folding energy of $\sim 130$ keV clearly improves the fit and better describes the spectrum. A broad ($FWHM \sim 2.8$ keV) iron emission line with centroid energy fixed at 6.4 keV was also needed in order to obtain acceptable fits. The iron line has an equivalent width of  53.5 eV. Finally, the addition of a narrow Gaussian component at $\sim$ 2 keV was also required in order to account for unphysical residuals in the XRT spectrum. To account for cross-calibration problems, a variable multiplicative constant for the PCA and HEXTE spectra (as compared to XRT) was added to the fits. $\chi_{\mathrm{red}}^2$ = 1.13 for 653 degrees of freedom (dof) was finally obtained. No additional reflection component is required to obtain a good fit. We derive an equivalent Hydrogen column value of $N_\mathrm{H} =(0.72_{-0.04}^{+0.01} ) \times 10^{22}$ atoms cm$^{-2}$. The fit for the XRT+PCA+HEXTE spectrum is shown in Fig. \ref{fig:1752_sp}. The spectral parameters obtained are listed in Table \ref{tab:spectral_par}. We also tried to fit the spectra using more sophisticated Comptonization models (\textit{comptt}, \textit{pexrav}) but the result was statistically worse than the obtained by using the model described above. \textit{comptt} and \textit{comptt}+reflection do not provide a valid fit and for \textit{pexrav} we obtain $\chi_{\mathrm{red}}^2$ = 1.26. for 650 dof.\\ 
The broad band spectrum of XTE J1752-223  is consistent with the source being a BHT in LHS. Its energy spectrum is dominated by a broken power-law component with photon indexes $\sim$ 1.2 and $\sim$ 1.5 and break energy $\sim$ 10 keV.  This is in agreement with the analysis performed by \cite{Wilms2006} on Cyg X-1 (see section \ref{discussion}). We also note that, in contrast to \cite{Shaposhnikov2009}, we do not need a disc component to fit the RXTE spectra. Only the addition of the XRT spectrum evidences that a disc component is necessary to achieve a good fit.\\

\begin{figure}
\centering
 \includegraphics[width= 8.8cm,height=6.5cm]{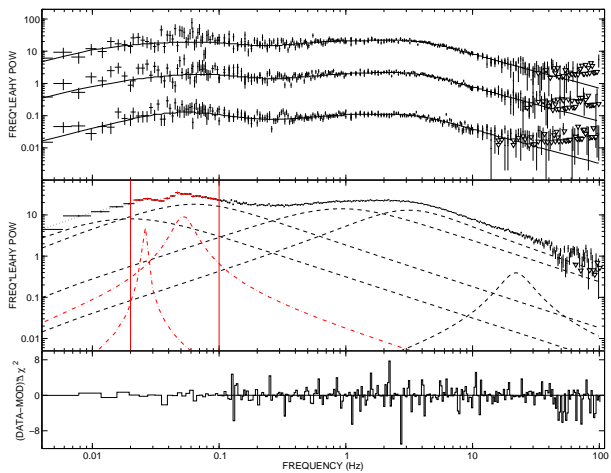}
\caption{PDS fits. Top panel: arbitrarily re-scaled PDS spectra for 3 different one-orbit observations (2/35, 16/35, 33/35) and their corresponding best fits (solid line). Triangles are $3\sigma$ upper limits.  Middle panel: fit of the average PDS. The four main components and the high frequency Lorentzian are shown in black, dashed lines. The (red) dotted-dashed lines correspond to the extra components added in the 0.02--0.1 Hz interval (see text). Bottom panel: residuals plotted as signed contribution to the chi square.}
\label{pds}
\end{figure}
\subsection{Power density spectrum}
Power density spectra (PDS) for each one-orbit pointing were computed using the same procedure outlined in \cite{Belloni2006}. We used stretches 512 s long  and energy channels STD2 0--31 (2--15 keV). PDS were fitted with \textsc{xspec} V.11. As for energy spectra we find that the shape of PDSs is almost constant during the whole observation, with it being possible to fit them by using 4 broad Lorentzians. Following \cite{Belloni2002}, one of this components is centred at zero frequency and it is kept frozen. To illustrate this, three one-orbit PDS (orbits 2, 16, 33) are shown in the upper panel of Fig. \ref{pds}. In these fits we notice the presence of residuals in the $\sim$ 0.02--0.1 Hz band. This extra component appears to vary between different orbits and can not be fitted by adding an extra narrow Lorentzian. In a second step, an average PDS was created (see lower panel in Fig. \ref{pds}). This PDS has a S/N ratio much higher than the previous ones, allowing a more accurate fit. We initially excluded the region  0.02--0.1 Hz and fitted with four Lorentzians. We find that these four broad components describe well the continuum of the average PDS. We obtain $\chi_{\mathrm{red}}^2$ =1.26 for 265 dof. Given the high signal-to-noise of the PDS it was not possible to fit all its wiggles and get a lower value for $\chi_{\mathrm{red}}^2$. In order to properly fit the 0.02--0.1 Hz region two extra components are required (dot-dashed lines in Fig. \ref{pds}). Moreover a weak, high frequency ($\sim 21$ Hz) component not visible in the one-orbit PDSs appears in the average spectra. In conclusion, seven Lorentzians are used, yielding $\chi_{\mathrm{red}}^2$ =1.28 for 281 dof (see lower panel in Fig. \ref{pds}). Table \ref{pdsfit} shows the parameters obtained from the fit. $eL_1$ and  $eL_2$ with $Q>2$ are probably related with weak quasi periodic oscillations (QPOs; see \citealt{Belloni2002}). The total rms in the 0.002--128 Hz band is $48.2\pm 0.1\%$.

\begin{table}
\begin{center}
\begin{tabular}{c|c|c|c}
\hline
Lorentzian 	&   Frequency (Hz)	&  Q  &  rms (\%)	\\
\hline
\hline
$L_0$	& $0^{+0.001}$ &      $0^{+0.28}$   &  $27.3 \pm 0.7$  \\
$L_1$	& $0.014\pm0.002$ &  $0.114\pm0.012$    &  $36.5 \pm 0.3$	 \\
$L_2$	& $0.07\pm0.01$ &  $0.04\pm0.01$ & $34.7 \pm 0.2$	\\			
$L_3$	&   $1.57\pm0.02$ &  $0.291\pm0.005$ &   $26.3 \pm 0.1$ \\
$L_4$	&  $21.1\pm1.9$ &  $1.7\pm0.5$ & $2.3 \pm 0.2$\\
\hline
$eL_1$	&   $0.0256_{-0.0006}^{+0.0002}$ & unconstrained  & $3.7 \pm 0.8$   \\
$eL_2$	&   $0.051\pm 0.002$ &  $2.7\pm0.6$ &  $8.9 \pm 0.6$   \\

\hline
\end{tabular}
\caption{Fit parameters for the broad Lorentzians, the high frequency Lorentzian ($L_{0,1,2,3,4}$; dashed line in Fig. \ref{pds}) and the extra ones added in the 0.02--0.1 Hz region ($eL_{1,2}$; dotted-dashed line in Fig. \ref{pds}).
}
\label{pdsfit}
\end{center}
\end{table}

\subsection{The spectrum of the fractional rms}
\begin{figure}
\includegraphics[width=8.5cm]{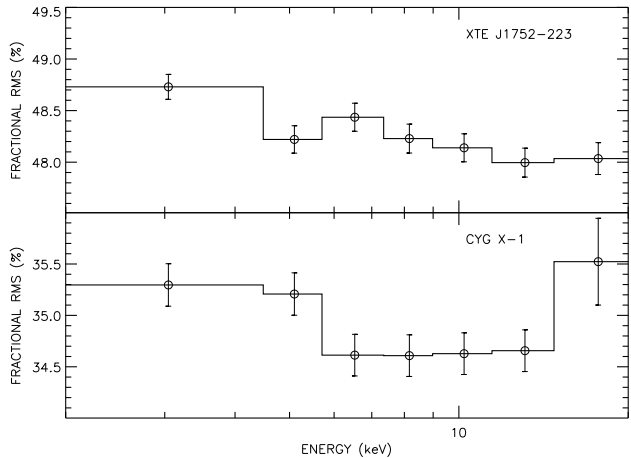}
\caption{Rms spectra for XTE J1752-223 and Cyg X-1 (0.002--128 Hz). 
\label{rmss}}
\end{figure}

In order to study the energy dependence of the rms we calculated it for seven energy bands within the range 2--20 keV. Average PDS corresponding to the STD2 channels  0--6 (2--4.5 keV), 7--9 (4.5--5.7 keV), 10--13 (5.7--7.3 keV), 14--17 (7.3--9 keV), 18--23 (9--11.4 keV), 24--31 (11.4--14.8 keV) and 32--45 (14.8-20.6 keV)  were used to compute the rms spectra (i.e. fractional rms vs. energy; see e.g. \citealt{Vaughan2003}, \citealt{Gierli'nski2005}). In Fig. \ref{rmss} we show our results for both XTE J1752-223 and Cyg X-1 within the 0.002--128 Hz band. We find that in both sources the fractional rms is almost independent of energy. We measure a rms of $\sim $48\% for XTE J1752-223 and $\sim 35$\% for Cyg X-1. This difference is consistent with the hardness-rms correlation generally observed in BHT (e.g. \citealt{Belloni2010}). We have also tried other frequency bands (e.g. 0.04--5, 10--128 Hz) always obtaining flat rms spectra.
 
\subsection{Time-lags}
\begin{figure}
\includegraphics[width=8.5cm]{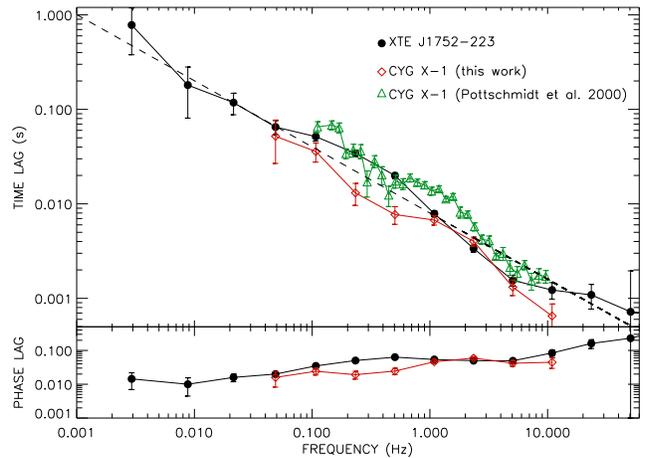}
\caption{Top panel: time-lag vs. frequency for XTE J1752-223 (black circles). For Cyg X-1 we over-plot the time-lags we find during the LHS observation we have analysed (open diamonds) and those found by  \citet{Pottschmidt2000} also during LHS (open triangles). The dashed line shows the relation $\Delta t \propto \nu^{-0.7}$ previously observed in Cyg X-1 (e.g. \citealt{Nowak1999}). Bottom panel: corresponding phase lags. 
\label{tlags}}
\end{figure}

We have computed time-lags  between soft and hard variability (see e.g. \citealt{Casella2004}). Following \cite{Pottschmidt2000}, we have used the energy ranges $\sim$ 2--4 keV and  $\sim$ 8--13 keV for the soft and hard bands, corresponding to the STD2 channels 0--5 and 15--27, respectively. Fast Fourier transforms of each band were computed and cross spectra were produced. Here, a positive lag means hard variability lagging soft variability. The obtained time-lags ($\Delta t$) as a function of the frequency ($\nu$) are shown in Fig. \ref{tlags} (black circles). Thanks to the quality of the data, we are able to cover the frequency range 0.003--30 Hz obtaining accurate time-lags determinations in most of the cases. Our results are in agreement with the relation ($\Delta t \propto \nu^{-0.7}$) observed in Cyg X-1 (e.g. \citealt{Nowak1999}; dashed line in Fig. \ref{tlags}). We find that time-lag drops from $\sim$ 0.4--1 s for 0.003 Hz to $\sim 0.0015$ s above 10 Hz. Evidences for a maximum delay are not found in the frequency range considered in this work. In Fig. \ref{tlags} we also show the time-lags obtained for Cyg X-1 by \cite{Pottschmidt2000} (open triangles). Although they only cover the range $\sim$ 0.1--10 Hz, the same trend is observed. We also include time-lags obtained from the Cyg X-1 observation analysed in this paper (open diamonds). The shape of the time-lag distribution is the same as the obtained by \cite{Pottschmidt2000}, but it is slightly re-scaled to smaller delays. This effect can be real or just be due to small changes in the energy bands used to compute the time-lags (e.g. changes in the sensitivity of the detectors). As shown in previous works (e.g. \citealt{Nowak1999}) time-lags increase when the difference between the energy bands considered does. For instance, if we use the energy range 14--45 keV (STD2 30--75) as the hard band, the higher delay observed for XTE J1752-223 is $\geq 0.6$~s.  

\section{Discussion} 
\label{discussion}  
We have performed a general analysis of XTE J1752-223 by looking at the spectral and timing properties of this recently discovered source. For a more complete understanding of the overall behaviour of the system  we have also made use of rms spectra and time-lags. We have analysed a very long RXTE observation which provides high S/N data and therefore an accurate determination of the different parameters has been possible.
\begin{itemize}
\item The combination of Swift (XRT) and RXTE (PCA+HEXTE) data have allowed us to  fit the energy distribution of XTE J1752-223 over the broad energy range 0.5-200 keV. The RXTE spectrum of the system is power-law dominated, the best fits being obtained by using a broken power-law model with photon indices $\Gamma_{1} \sim 1.5$ and $\Gamma_{2} \sim 1.2$. It is remarkable that this is also the model which best reproduce the energy spectra of the canonical BHB Cyg X-1 (\citealt{Wilms2006}). These authors monitored the source during five years finding interesting correlations between  $\Gamma_{1}$ and $\Gamma_{2}$ but also between ($\Gamma_{1}-\Gamma_{2}$) and $\Gamma_{1}$. The values we found for XTE J1752-223 lie on both correlations but correspond to a state slightly harder than the hardest reported by \cite{Wilms2006}. On the other hand, the difference between spectral indices ($\Gamma_{1} - \Gamma_{2} \sim 0.23$) and high energy cut-off ($\sim$ 133 keV)  are in agreement with those measured for Cyg X-1 during LHS. The latter is also consistent with values found in classical BHT like GX 339-4 (\citealt{Motta2009}) or GRO J1655-40 (\citealt{Joinet2008}) during LHS.\\
The RXTE spectrum of XTE J1752-223 does not require of a disc component to get a good fit, but the disc becomes evident when adding the Swift data. Our analysis reveals the presence of a cold disc with an inner radius temperature of $\sim 0.3$ keV. From  the disc black-body normalization component (see \citealt{Mitsuda1984}) it is possible to derive the value of the inner radius of the accretion disc. In particular, assuming a distance in the range 2-8 kpc, a $10\sol$ BH and an inclination $ \leq 70^{\circ}$ (eclipses have not been observed) we find an inner disc radius in the range 9--43 gravitational radii. Using the same distance interval we derive a luminosity (0.5-200 keV) of $2.5-40\times10^{36}$ erg s$^{-1}$. 

\item The continuum of the PDS of XTE J752-322 can be described by four broad Lorentzians ($L_{0,1,2,3}$) with rms above $25\%$, a high frequency Lorentzian ($L_4$) and two QPO-like features ($eL_{1,2}$). This is very similar to the found on Cyg X-1 by \cite{Pottschmidt2003} during some of the LHS epochs they analyse. Associating $L_0$ with the low frequency power-law component they used, one is tempted to match the other components ($L_{1,2,3,4}$)  one-to-one. For Cyg X-1 the Lorentzians have lower rms values and higher frequencies, which is consistent with XTE J752-322 being in a harder state. We note that in both cases $L_{1,2,3}$ show large rms values whereas the high frequency Lorentzian ($L_4)$  present in both studies is much weaker. On the other hand, frequency ratios are different. \cite{Pottschmidt2003} report $\nu_2/\nu_1\sim$~6--9, $\nu_3/\nu_1\sim$~25--35, $\nu_4/\nu_1\sim$~80--200, $\nu_3/\nu_2\sim$~3--4, $\nu_4/\nu_2\sim$~11--25, $\nu_4/\nu_3\sim$~4--6 whereas we find $5.2\pm0.9$, $110\pm11$, $1483\pm198$, $21\pm3$, $285\pm48$ and $13\pm1$, respectively. The ratios might be affected by the different definition of $L_{0}$ adopted in both works (Lorentzian/power-law). We have tried to fit our average PDS with a power-law, which results very flat, instead of $L_{0}$. Significant variations on frequency are only observed for $L_1$ (i.e. $\nu_1$). We also note that weak QPOs like the ones we detect in the 0.02--0.1 Hz band (i.e. $eL_{1,2}$) are also detected by \cite{Pottschmidt2003} in Cyg X-1. Again, their frequencies are slightly higher (0.1--1 Hz) than those we observe in XTE J1752-223.

\item Rms spectra tell us how variability depends on the energy band considered. Their shape and comparison with energy spectra also provides clues on the origin of the observed variability (e.g. \citealt{Wilkinson2009}). Previous works have studied the shape of rms spectra during different states in several BHT (see e.g. \citealt{Gierli'nski2005} for a general description). In particular, two different rms spectral shapes have been observed during LHS: flat (e.g. XTE J1550-564) and smoothly decreasing with energy (e.g. XTE J1650-500). In this paper we show that the 2-20 keV rms spectrum of XTE J1752-223 is flat within 1\%. The same behaviour is observed in the LHS observation of Cyg X-1 that we have analysed. This is expected if the variability is produced by changes in the normalization of the entire spectrum, but keeping the spectral shape constant. For the case of XTE J1752-223 (and Cyg X-1), where no disc component is present in the RXTE spectrum,  our results are consistent with variability being due by variations in the normalization of the Comptonization (power-law) component. \cite{Gierli'nski2005} also discuss possible features caused by reflection components in the rms spectrum. For the case in which reflection and continuum components are not correlated, they predict the presence of $\sim 2-5\%$ absorption features in the rms spectra. These features are not detected in our rms spectrum. We note that this is consistent with our spectral fitting (it does not require a reflection component), although the broken power-law shape mimics the effect of reflection (\citealt{Wilms2006}).  
 
\item Time-lags between soft and hard photons are expected in Comptonized spectra as a result of the different number of scatterings that they undergo. In this framework, one expect hard variability delayed relative to soft variability. This is observed in XTE J1752-223, being the time-lag a function of the frequency. We find that our measurements are well described by $\Delta t \propto \nu^{-0.7}$. However, we note that this is only a qualitative description of the time-lags since clear deviations are observed. Clear flattenings suggesting a maximum or minimum delay are not observed in our data, but above 10 Hz our results are consistent with $\Delta t\sim0.0015$~s. This is in agreement with what it has been observed in Cyg X-1 (e.g. \citealt{Nowak1999}, \citealt{Pottschmidt2000}). The highest delay we observe is $\gtrsim 0.5$ s. However, our frequency coverage extends down to 0.003 Hz, and our highest lag is one order of magnitude larger that the highest found by the cited works ($\sim 0.05$~s at 0.1 Hz). As noticed by \citet{Nowak1999a} a $\sim 0.05$ s delay is difficult to be reproduced by either Comptonization models, sound speed propagation or gravitational free fall, unless the length scales involved were very large ($\sim 10^3$ gravitational radii). For XTE J1752-223  the highest delay ($\gtrsim 0.5$ s) yields a characteristic length of $\sim 10^4$ gravitational radii and there is no evidence to discard higher lags at frequencies lower than 0.003 Hz. These large lags require of either, those long length scales or low propagation velocities ($< 0.001c$) difficult to be reproduced by the current theoretical models (e.g. ADAF).
Alternative scenarios show that time-lags are expected for pivoting power law emission (\citealt{Kording2004}). \cite{Kazanas1997} and \cite{Pottschmidt2000} suggest that they can be related with the accretion disc or outflows, respectively. Our results favour these latter scenarios in which time-lags are not purely created by Comptonization processes within the corona.               
\end{itemize}
\section{Conclusions}
An unusually long RXTE observation of the X-ray transient XTE J1752-223 is presented in this paper. The quality of the data have allowed us to obtain high S/N energy spectrum, PDS, rms spectrum and time-lags for this new source. All the obtained results are consistent with a black hole binary in the hard state. In particular, we find a behaviour similar to that exhibited by Cyg X-1 during hard state, but XTE J1752-223 happens to be in a slightly harder state. However, we note that there are two important differences between these two systems: Cyg X-1 is so far a persistent black hole binary and it harbours a high-mass companion. XTE J1752-223 is a transient and probably harbours a low-mass donor. Future multi-wavelength campaigns will probably provide new clues to the fundamental properties of this new black hole candidate. 
\vspace{1cm}

\noindent The research leading to these results has received funding from the European Community's Seventh Framework Programme (FP7/2007-2013) under grant agreement number ITN 215212 \textquotedblleft Black Hole Universe\textquotedblright. SM and TB acknowledge support to the ASI grant I/088/06/0. TMD, SM, TB and  DP acknowledge hospitality during their visits to IUCAA (Pune).

\bibliographystyle{mn2e.bst}
\bibliography{RX.bib} 

\label{lastpage}
\end{document}